\documentclass[preprint, sort&compress, number]{elsarticle}

\usepackage[T1]{fontenc}
\usepackage{amsmath}
\usepackage{graphicx}

\begin{document}
 
\begin{frontmatter}

\title{A generalized scheme for characterizing orientational correlations in condensed phases of high symmetry molecules: SF$_6$ and C$_{60}$}

\author[1]{L\'{a}szl\'{o} Temleitner \corref{cor1}}
\ead{temleitner.laszlo@wigner.hu}

\cortext[cor1]{Corresponding author}

\affiliation[1]{organization={Institute for Solid State Physics and Optics, Wigner Research Centre for Physics}, addressline={Konkoly Thege {\'u}t 29-33.}, postcode={1121}, city={Budapest}, country={Hungary}}





\def\beq{\begin{equation}}
\def\eeq{\end{equation}}


\date{}



\begin{abstract}

The orientational correlation scheme introduced earlier for tetrahedral molecules is extended for being able to classify orientational correlations between pairs of high symmetry molecules. While in the original algorithm a given orientation of a pair of tetrahedral molecules is characterized unambiguously by the number of ligand atoms that can be found between two planes that contain each centre and perpendicular to the centre-centre connecting line, in the generalized algorithm, the planes are replaced by cones, whose apex angles are set according to the symmetry of each molecule. To demonstrate the applicability of the method, the octahedral-shaped SF$_6$ molecule is studied in a wide range of phases (gaseous, supercritical fluid, liquid and plastic crystalline) using classical molecular dynamics. By analyzing the orientational correlations, a close-contact region in the first coordination shell and a medium-range order behaviour are identified in the non-crystalline phases. While the former is invariant to changes of the density, the latter showed longer-ranged correlations as density is raised. In the plastic crystalline state, fluorine atoms are oriented along the lattice directions with higher probability. To test the method for icosahedral symmetries, the crystalline structures of room temperature C$_{60}$ is generated by three sets of potentials that produce different local arrangements. The novel analysis provided quantitative result on preferred arrangements. Submitted version of the open-access article [Journal of Molecular Liquids, 2021, https://doi.org/10.1016/j.molliq.2021.116916]

\end{abstract}

\begin{graphicalabstract}
\resizebox{0.96\textwidth}{!}{\includegraphics{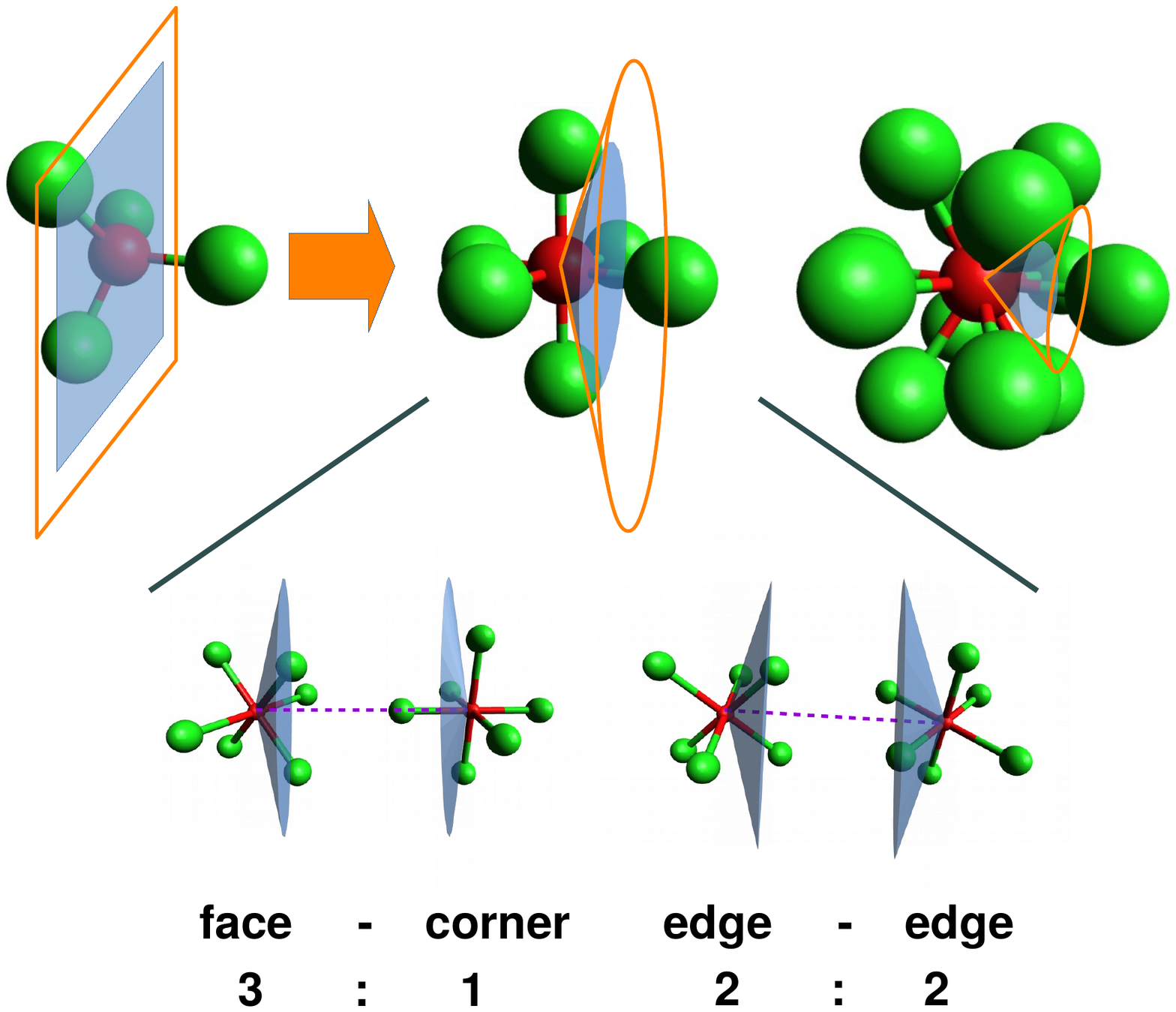}}
\end{graphicalabstract}





\end{frontmatter}

\section{Introduction}

A most important issue in the structure analysis of disordered materials is to find suitable and simple representation(s) of the studied material that can emphasize its characteristic property. Considering orientational correlations in condensed phases of high-symmetry molecules (including the ones whose shape belong to the most symmetrical ones called 'Platonic solids': tetrahedron, octahedron, cube, icosahedron and dodecahedron), there is a lack of suitable descriptions: atomic radial distribution function (RDF, g(r)) analysis is a frequently applied method, but it does not take into account the symmetry of molecules in a direct way. On the other hand, the application of symmetry adapted functions is restricted mainly to crystalline phases. In 2007 Rey\cite{rey_2007} introduced a scheme that classifies any mutual arrangement between pairs of tetrahedral molecules to an ideal arrangement. In order to characterize a given arrangement, two parallel planes are taken, each contains one of the molecular centres and perpendicular to the line connecting them. Then, the classification is performed by the number of corner (or ligand) atoms of each molecule are between the two planes. This way, six different kinds of ideal arrangements have been defined, covering all possible orientations, as corner to corner(1:1), corner to edge(1:2), corner to face(1:3), edge to edge(2:2), edge to face(2:3) and face to face(3:3). Thus, the probability of each arrangement can be determined for any centre-centre distances. The advantage of this method that it provides quantitative results and represents correlations continuously in distance rather than providing an average over a coordination shell. The method has contributed to the elucidation of the structure of liquids consisting of regular\cite{rey_2009, pothoczki_2009, morita_2009, pothoczki_2009b, pothoczki_2015} and distorted tetrahedral shapes of molecules\cite{pothoczki_2010, pothoczki_2017} and recently, of ionic conductors\cite{lacivita_2018}.

Until now, this method has not been used for pairs of molecules whose shapes are non-tetrahedral Platonic. Note using the parallel planes as characterizing objects would provide useless results: only one class is formed as the non-tetrahedral Platonic solids possess inversion symmetry, which assigns exactly half of the corner atoms oriented to the other molecule in almost any case. Thus, the method should be generalized in another way: by keeping each class (corner, edge and face) and introducing orientation cones instead of planes as discussed in Section \ref{rey_general}. To demonstrate the applicability of this approach, orientational correlation analysis in condensed phases of the octahedral-shaped sulfur-hexafluoride (SF$_6$) and room temperature crystalline phase of the truncated icosahedral-shaped C$_{60}$ (based on various potential models) are selected.

Sulfur hexafluoride (SF$_6$) is regularly used as an insulator in high-voltage circuit breakers\cite{nosseir_1965}. Apart from the technical importance, the high symmetry of the molecules and a peculiar body-centered plastic crystalline phase attracted much attention from a scientific point of view\cite{taylor_1976, dolling_1979, pawley_1981, dove_1983, dove_1984, dove_2002}. This phase is stable between the point of sublimation at 223K and phase transition to the low-temperature monoclinic phase\cite{cockroft_1988} occurs at 96K under ambient pressure\cite{dolling_1979}. While the centre of mass of the molecule possesses translational symmetry, apparent molecular rotations are only hindered by neighbouring molecules. Neutron diffraction\cite{taylor_1976, dolling_1979, cockroft_1988} and molecular dynamics (MD) modelling\cite{dove_1983, dove_1984} studies usually discussed correlations relative to the ordered crystalline lattice and concluded that fluorine atoms oriented close to the cubic cell axis. While the orientation of the molecule relative to the lattice is frequently taken into account in MD studies addressing the low-temperature phase\cite{pawley_1983} and the ordered-plastic phase transitions\cite{torchet_1990, boutin_1994}, there are only a few MD\cite{dove_1983, dove_1984}, and one\cite{dove_2002} Reverse Monte Carlo (RMC) studies that consider local atomic arrangements in terms of atomic RDF-s. RMC was developed originally for liquids and amorphous materials\cite{mcgreevy_1988} and provide sets of atomic configurations consistent with measured diffraction data without the need to use any potential.

The gaseous, liquid and supercritical fluid states have triggered only a few diffraction studies\cite{powles_1983, strauss_1994}. MD simulations are compared to experimental properties of these phases by Dellis and Samios\cite{dellis_2010}, using potential sets optimized for describing supercritical fluid\cite{strauss_1994} and crystalline phases\cite{pawley_1981, kinney_1996}. Strauss et al.\cite{strauss_1994} used RMC modelling, and their description was focused on atomic RDFs. According to my best knowledge, there has not been any investigation performed yet that would take into account the symmetry of the SF$_6$ molecule and aim at mutual orientations both in liquid and crystalline phases.

C$_{60}$ (fullerene, buckyball) has numerous phases depending on temperature and pressure\cite{moret_2005, alvarez-murga_2015}. At low-temperature and ambient pressure, it forms a simple cubic molecular crystal with some disorder. At 260K molecules start to rotate, thus inducing a phase transition\cite{david_1992} to the orientationally disordered face-centered cubic structure, stable at and above room temperature. Early diffraction studies\cite{li_1991, soper_1992, leclercq_1993} clarified the structure of the molecule: they observed a difference between single and double bonds ('D') separating pentagon ('P')-hexagon ('H') and hexagon-hexagon faces, respectively. Concerning the room temperature phase, powder diffraction and RDF (as derived from powder diffraction) studies suggested free and independent rotation\cite{david_1992} and a lack of intermolecular ordering\cite{soper_1992}. On the other hand, single-crystal studies provided evidence of a slightly (within 20\% intensity modulation) preferred orientation of molecules\cite{chow_1992} with more pentagonal than hexagonal faces turning towards nearest neighbour directions\cite{schiebel_1996}, and of local ordering of molecules by diffuse scattering\cite{moret_1992, launois_1995, chaplot_1995, pintschovius_1995a}.

To describe the structure and take into account orientational ordering, various potential sets have been developed, as reviewed by Launois et. al\cite{launois_1997, launois_1999} and Chaplot et. al\cite{chaplot_2001}. These potential sets spread from simple van der Waals interactions on carbons\cite{girifalco_1992, monticelli_2012}, additional van der Waals centres on double-bonds with charges\cite{sprik_1992b}, additional charges on single and double-bonds\cite{lu_1992}, split charge on double-bonds\cite{pintschovius_1995c, chaplot_1995}, anisotropic van der Waals\cite{kita_2006}, mean-field\cite{michel_1997} to ab-initio\cite{tournus_2005}. The main reason to overcome the simple van der Waals potential is that it does not produce the experimentally observed low-temperature phase\cite{quo_1991}. Among the reviewed forcefields, none of them is able to describe all of the observed properties\cite{launois_1999}.
Orientational correlations, for instance, depend significantly on the applied forcefield\cite{chaplot_2001}

Details of the MD simulations and calculation of the orientational correlations are presented in Section \ref{methods}. Section \ref{resdis} contains the multiphase (gaseous, liquid, fluid, plastic crystalline) investigation of orientational correlations of octahedral SF$_6$ using the extended method. It is followed by the quantitative analysis of short-range order by different forcefields of room temperature C$_{60}$, consisting of truncated icosahedral molecules. Finally Section \ref{conclusions} summarizes the results.

\section{\label{rey_general}Theory}

\subsection{Generalization of the classification scheme} 

To extend the classification scheme introduced by Rey\cite{rey_2007} originally for tetrahedral molecules, the question of how many corner atoms of each molecule is between two \emph{planes} \emph{perpendicular} to the connecting line of the molecular centres should be replaced. Instead of \emph{perpendicular planes}, we should consider \emph{orientational cones} with proper \emph{angle} constructed by supposing:
\begin{enumerate}[1.]
 \item The directions of corner atoms from the centre are distributed evenly on the surface of a sphere. 
 \item The average number of corner atoms from one molecule inside the cone should be 2 when the molecule is oriented randomly.
 \item Only 1, 2, or 3 corner atoms are allowed to be inside the cone.
\end{enumerate}

The first two conditions determine the half apex angle ($\gamma$) of the cone that belong to the molecule possessing $N_{corner}$ corner atoms. According to them, the ratio of the solid angle of the orientation cone and the total solid angle should be equal to the ratio of the average number of corner atoms inside the cone at random orinetation of the molecule and $N_{corner}$:
\begin{equation}\label{eq:gamma_basic}
 \frac{2}{N_{corner}} = \frac{2\pi \int_{\cos{\gamma}}^1 d (\cos{z})}{4\pi} = \frac{1-\cos{\gamma}}{2},
\end{equation}
after some re-arrangement:
\begin{equation}\label{eq:gamma}
 \gamma = \arccos{\left ( 1 -\frac{4}{N_{corner}}\right ) },
\end{equation}
is obtained. This corresponds to the definition of a plane in the case of regular or slightly distorted tetrahedral molecules.

The next task is the determination of the random orientation probabilities for each situation: the third condition limits them to three cases only. However, the second condition fixes the average to be 2, so the $P_i$ probabilities of having $i$ number of corner atoms inside a randomly directed cone for one and three atoms should be equal. Thus, only a single $P_i$ probability parameter exactly determines the other two:

\begin{equation}\label{eq:general_rule}
 P_1=P_3=\frac{1-P_2}{2}
\end{equation}

The derivation and calculation method of $P_i$ with the help of corner-centre-corner bond angles is presented in SI 1.

Combining these probabilities for pairs of molecules according to the appendix of \cite{rey_2007}, the asymptotic values of each of the 6 groups can be obtained for molecules having tetrahedral, octahedral and icosahedral symmetries (see TABLE \ref{tab:reygrprob}).

\begin{table}[ht]
\caption{\label{tab:reygrprob} Asymptotic probabilities of each groups for different symmetry of molecules obtained by taking into account the $p_{1:1}=p_{3:3}=P_1^2$, $p_{1:2}=p_{2:3}=2P_1P_2$, $p_{2:2}=P_2^2$ and $p_{1:3}=2P_1^2$ relations for single component molecular pairs (according to \cite{rey_2007}). $P_1$ is the probability to find one corner atom inside the cone, which is determined on the basis of angular separation of the number of corner atoms ($N_{corner}$) and half apex angle of the cone ($\gamma$).}
\resizebox{\textwidth}{!}{
\begin{tabular}[c]{cccccccccc}
\hline
molecule & $N_{corner}$ & $\cos{\gamma}$ & $\gamma$ & $P_1=P_3$ & $P_2=1-2P_1$ & $p_{1:1};p_{3:3}$ & $p_{1:2};p_{2:3}$ & $p_{2:2}$ & $p_{1:3}$ \\
symmetry & & & [$^o$] & & & [\%] & [\%] & [\%] & [\%] \\
\hline
tetrahedral & 4 & 0 & 90 & 0.17548 & 0.64904 & 3.079 & 22.779 & 42.125 & 6.159 \\
octahedral & 6 & $\frac{1}{3}$ & 70.53 & 0.22075 & 0.55850 & 4.873 & 24.658 & 31.193 & 9.746 \\
icosahedral & 12 & $\frac{2}{3}$ & 48.19 & 0.25251 & 0.49498 & 6.376 & 24.997 & 24.501 & 12.752 \\
\hline
\end{tabular}
}
\end{table}

The discussion of the studied system using this formalism is facilitated by the fact that each of the 6 ideal arrangement is broadly used (for instance octahedral molecules the figure 1 of REF \cite{dove_1983} contains almost all ideal arrangement except the edge to corner) and a given arrangement of a molecular pair is categorized unambiguously into one of the 6 idealized ones.

\section{Computational details}\label{methods}

\subsection{Simulations}\label{sim}

Classical MD simulations with flexible molecules have been performed by the GROMACS\cite{gromacs_2015} package using version 2016.3. All simulations used Lennard-Jones interaction parameters and partial charges to model intermolecular and distant intramolecular dispersive and Coulombic interactions. Except for the case of one forcefield where the molecules are kept rigid, intramolecular interactions between adjacent atoms are modelled with harmonic bond stretching, angle bending and improper dihedral potentials.

Among the \emph{sulfur-hexafluoride} forcefields reviewed by Dellis and Samios in table 1 of REF. \cite{dellis_2010}, both optimized (referred by the '-opt' postfix) and original forms of Pawley\cite{pawley_1981}, Kinney\cite{kinney_1996}, Strauss\cite{strauss_1994} and 7Sites\cite{dellis_2010} (referred as 'Pawley', 'Kinney', 'Strauss' and 'Dellis', respectively) are used in this study in flexible form\cite{dellis_2010}.

Concerning the case of room temperature phase of \emph{fullerene}, three potential sets have been used to model intermolecular interactions: the 'Girifalco'\cite{girifalco_1992} forcefield with Lennard-Jones potential on each carbon atom; the 'Lu'\cite{lu_1992} forcefield with modified Lennard-Jones parameters on the atoms and additional charges on single and double-bonds; and the 'Sprik'\cite{sprik_1992b} forcefield with charges and Lennard-Jones potentials on each carbon atom and double-bonds, with Lorentz-Berthelot combination rules. According to 'Sprik', molecules were kept rigid, while in the former two forcefields the flexible potential set of Monticelli et al.\cite{monticelli_2012} has been used.

Further details are presented in SI 2.

\subsection{Calculating orientational correlations}

In the saved trajectories of SF$_6$, positions of sulfur and fluorine served as the central and corner atoms, respectively.

For the C$_{60}$ particle configurations the icosahedral symmetry was reconstructed for each molecule. One of the possible generation of the C$_{60}$ molecule from a perfect icosahedron is to truncate each edge at a given ratio around each corner. As in a \emph{corner} of the icosahedron 5 edges join, a \emph{pentagonal face} is formed after the truncation. On the other hand, each triangular \emph{face} of the icosahedron is truncated and a \emph{hexagonal face} is created. Then, the non-truncated \emph{edge}s of the icosahedron are the \emph{double bond}s of C$_{60}$ where two hexagonal faces join. As in the following, non-endohedral fullerenes are taken, the centres of them were calculated as the mean of atomic coordinates and the centre of each pentagon served as the virtual corner atom of the icosahedron at calculations of centre-centre RDF and orientations.

Following the symmetry reconstruction, the categorization of the orientations of each pair of molecules into the 6 categories as a function of centre-centre distances is performed using custom-written software. During the examination of the pairs, the numbers of corner atoms inside the cones centred on the centre-centre connecting line and having half apex angles of 70.53$^o$ for SF$_6$ and 48.19$^o$ for icosahedral representant of C$_{60}$ have been taken into account, according to the requirement of section \ref{rey_general}. Next, each contribution in a given distance bin is normalized by the sum of them. Finally, the probabilities calculated from each of the 101 configurations are averaged.

\section{Results and discussion\label{resdis}}

\subsection{Orientational correlations in condensed phases of sulfur hexafluoride}

Neutron diffraction measurements at various thermodynamical conditions (published for gaseous\cite{powles_1983}: 293.15 K and 10.2 bar, $\rho$=0.0021 {}\AA $^{-3}$; supercritical fluid\cite{strauss_1994}: 398 K; at 128 bar, $\rho$=0.0245 {}\AA $^{-3}$; at 155 bar, $\rho$=0.0289 {}\AA $^{-3}$; at 394 bar, $\rho$=0.0404 {}\AA $^{-3}$; at 1827 bar, $\rho$=0.0534 {}\AA $^{-3}$ and plastic crystalline\cite{dove_2002, rmcprofile_tutorial} states: 190 K and 1 bar, $\rho$=0.0685 {}\AA $^{-3}$) allow us to observe the changes of orientational correlations in the framework of the modified algorithm. Similarly, a large variety of forcefields are available for this material. In SI. 4., 
these forcefields are tested if they are able to reproduce the available total scattering neutron diffraction data (calculation method of them are presented in SI. 3.) 
and density. Concerning these quantities, the 'Dellis' ('7sites' in\cite{dellis_2010}) forcefield of Dellis and Samios\cite{dellis_2010} was proven to be the best and provided a good agreement with experimental results. Note that this forcefield had been optimized originally against densities in the liquid and supercritical fluid regions, it also performed well close to the triple point (225 K and 10.1 bar, $\rho$=0.0532 {}\AA $^{-3}$) and in the plastic crystal phase that had not been parts of the original range of validation. Thus, in the following, results of MD simulations from this potential set are used. 

\begin{figure}[htp]
 \begin{center}
  \rotatebox{0}{\resizebox{0.8\textwidth}{!}{\includegraphics{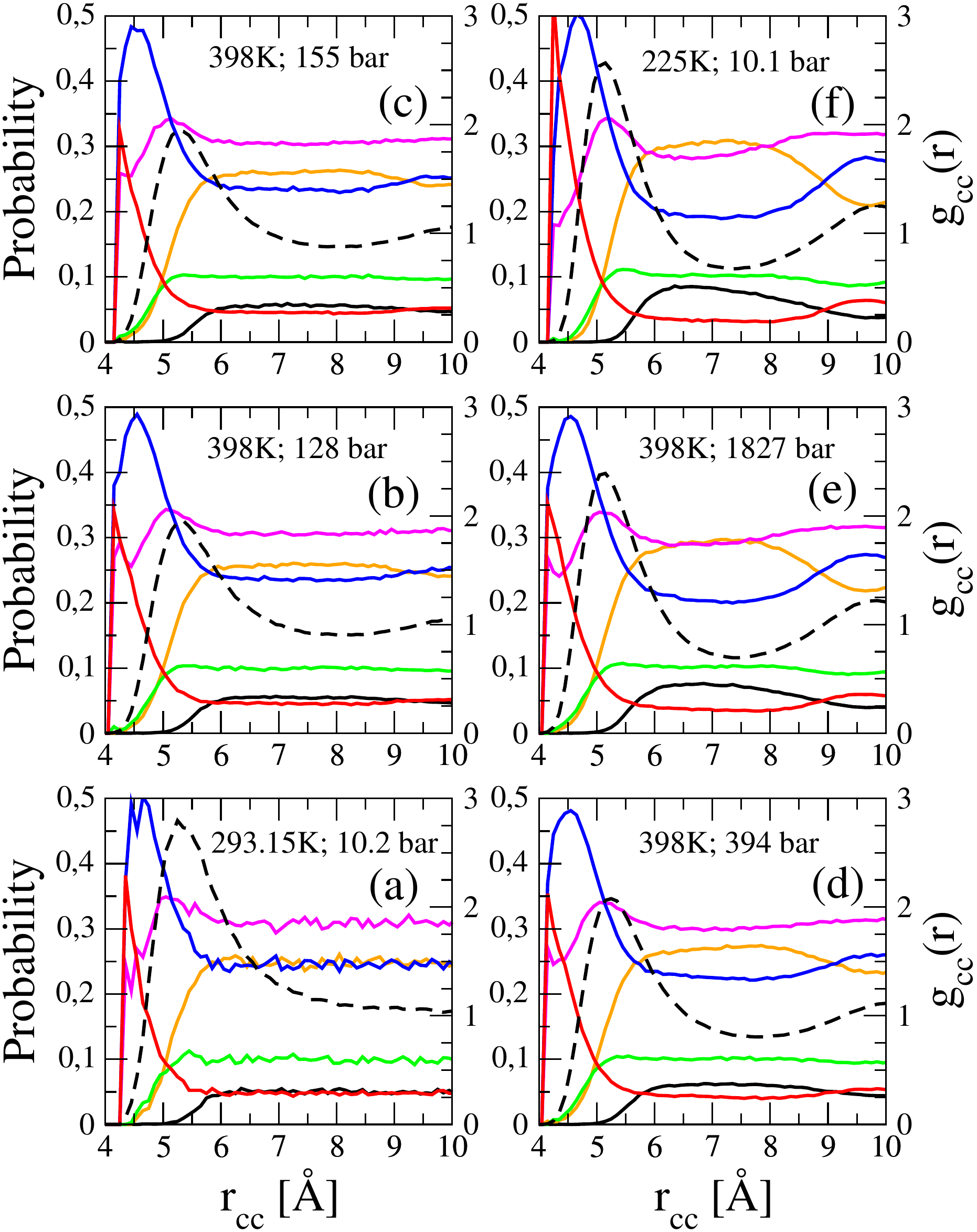}}}
  \caption{\label{fig:sf6_reysh} Orientational correlation probabilities focusing on short-range at 6 non-crystalline states of SF$_6$ using the classification method of Rey\cite{rey_2007} (extended here for other Platonic shapes). Probability scales are on the left, while scale for the molecular centre-centre radial distribution function (black dashed lines) on the right side of each subfigure. Face-face (3:3): red lines; corner-corner (1:1): straight black lines; edge-face (2:3): blue lines; corner-edge (1:2): orange lines; corner-face (1:3): green lines; edge-edge (2:2): magenta lines.}
 \end{center}
\end{figure}

\begin{figure}[htp]
 \begin{center}
  \rotatebox{0}{\resizebox{0.8\textwidth}{!}{\includegraphics{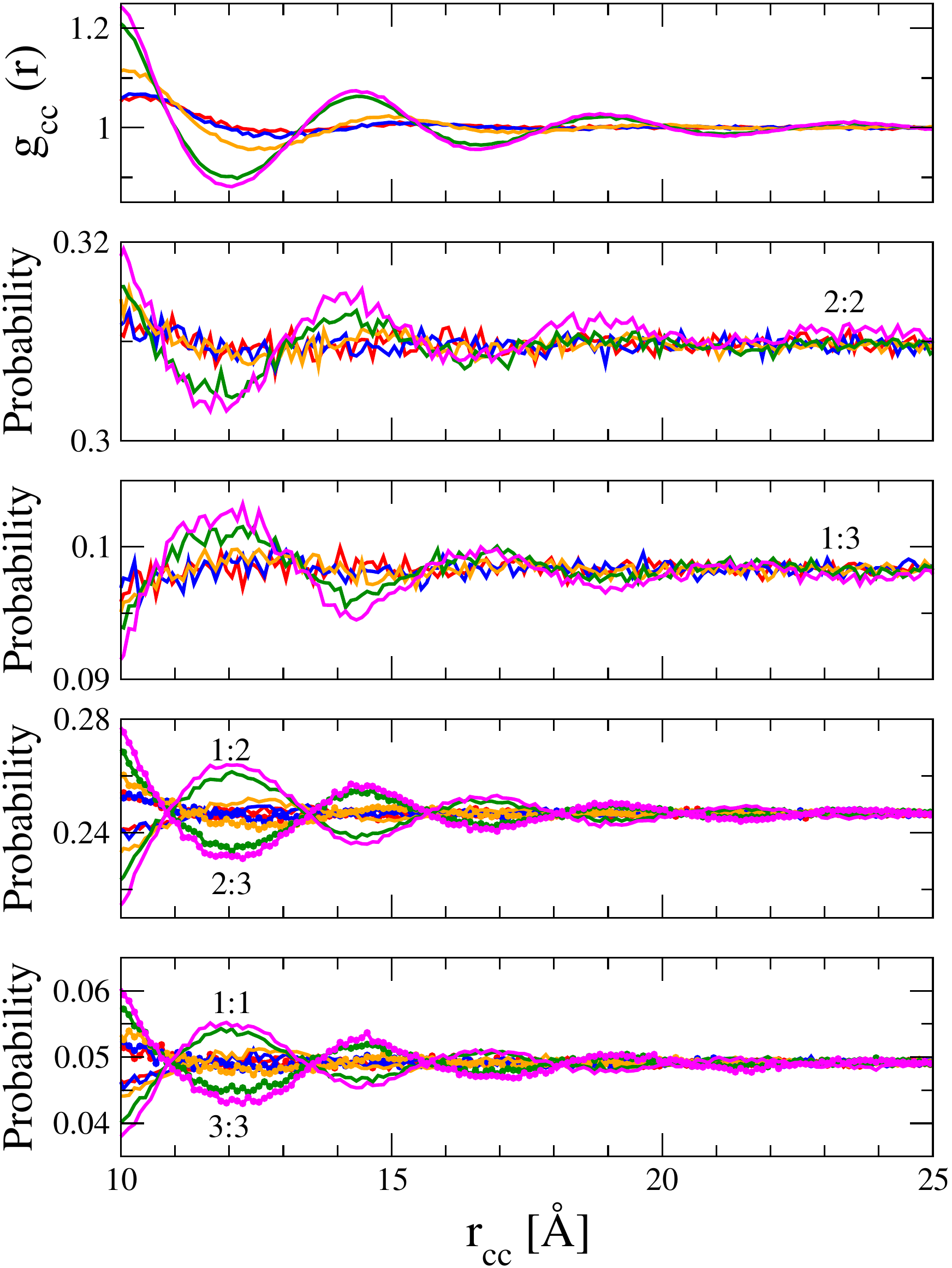}}}
  \caption{\label{fig:sf6_reylong} The molecular centre-centre radial distribution function (on top) and orientational correlation probabilities (the lower 4 subfigures) focusing on the 10 to 25{}\AA{} range at 5 non-crystalline states of SF$_6$ using the classification method of Rey\cite{rey_2007} (extended here for other Platonic shapes). 398K (supercritical states), 128 bar: red lines; 155 bar: blue lines; 394 bar: orange lines; 1827 bar: dark green lines; 225K, 10.1 bar (liquid): magenta lines. In the bottom two figures, the groups of 1:1 and 1:2 represented by straight lines, while for 3:3 and 2:3 by straight lines with circles.}
 \end{center}
\end{figure}

FIGURES \ref{fig:sf6_reysh} and \ref{fig:sf6_reylong}
contain the probabilities of each group with the molecular centre-centre radial distribution function in 6 non-crystalline states: one gaseous (FIGURE \ref{fig:sf6_reysh}a), 4 supercritical fluid (FIGURE \ref{fig:sf6_reysh}a-e) and a close-to-triple point liquid (FIGURE \ref{fig:sf6_reysh}f).

In order to obtain a general picture, we start with the examination of the gaseous state (FIGURE \ref{fig:sf6_reysh}a). In this phase the colliding molecules determine the behaviour of the system, so we expect that the direct contact of atoms characterizes the short-lived orientational correlations. The molecular centre-centre RDF reflects this feature as only a broad first maximum is present with a maximum at 5.3 {}\AA{}, which slowly decays at higher distances. Focusing on the orientational correlations, they definitely depend on the molecular centre-centre distance: the face-face (3:3) arrangement is frequent at the shortest distances, while farther away, the probability drops until it reaches its equilibrium value at 6{ }\AA{}, slightly over the first maximum of the centre-centre RDF. Next, the edge-face (2:3) and further, the edge-edge (2:2) arrangement probabilities rise and have definite maxima at distances shorter than the maximum of the centre-centre RDF. The remaining corner-face (1:3), corner-edge (1:2) and corner-corner (1:1) arrangements have only the rising part in probability without a characteristic maximum: they reach their asymptotic values roughly at 6\AA{}.

The observed behaviour has a simple geometric reason: the centre to a centre of a face, to mid of edge and to the corner atom distances are increasing, respectively. If we wish to pack atoms of molecular pairs as close as possible, the order of appearance of the arrangements should be 3:3, 3:2, 2:2 and 1:3, 1:2 and 1:1 as the centre-centre distance increases. A similar tendency is valid in most of the tetrahedral molecules\cite{rey_2009} when this method is applied. However, in the present case, there are apparent differences from it, namely the intensity of 3:3 is not 1 at the shortest distances, but it is somewhat comparable to the probability of 2:3 arrangements. The explaination may be that there are only a few pairs in this region.

Extending the discussion to the supercritical fluid and liquid states (FIGURE \ref{fig:sf6_reysh}a-f), we can conclude that up to about 5.5 {}\AA {} orientational correlations do not alter in comparison with the gaseous state. This common behaviour is the direct consequence of the close contact of corner atoms as discussed above.

While the gaseous state does not show any significant correlations, neither in the centre-centre, nor orientations beyond the first coordination shell, the centre-centre RDF becomes more structured as the density of the system increases. Oscillations also appear (FIGURES \ref{fig:sf6_reysh} and \ref{fig:sf6_reylong}) in the orientational correlations above 5.5..6 {}\AA{}, which is slightly less than the upper limit of the first coordination shell. These oscillations are observed up to the upper limit of the second coordination shell in the two lowest density fluid states, up to the third maximum of centre-centre RDF for 394 bar and up to the upper limit of the fourth shell for the highest density states. It is important to note here that the simulated molecules were flexible and atomic coordinates are used without any further alignment during the calculation of orientational correlations, which slightly biased the asymptotic probability value of some groups (the largest deviation is about 0.002 for the 2:2 pairs).

Concerning the different kind of arrangement at medium-range, they do not show any preference for a specific correlation, but seemingly the probabilities of the 3:3 and the 3:2 arrangements are in opposite phase with probabilities of the 1:1 and the 1:2. Importantly, except the first common node point of probabilities of the 3:3 with the 1:1, of the 2:3 with the 1:2 and where the centre-centre RDF passing through unity, they coincide with each other. Moreover, the probability of arrangements having more corner atoms inside the orientational cones than the average 4 (the 3:3 and the 3:2) are in phase with, while arrangements indicate fewer corner atoms are in opposite phase with the centre-centre RDF.

This observed medium-range order suggests a relationship with the tetrahedral-shaped CCl$_4$ liquid\cite{rey_2007}, where similar behaviour was found. However, due to the remarkable short-range order correlations, the medium-range order starts there about from the third coordination shell and correlations are slightly shifted in phase with the centre-centre RDF (see FIGURES 4 and 5 of\cite{rey_2007}). Thus, the two systems behave differently.

\begin{figure}[htp]
 \begin{center}
  \rotatebox{0}{\resizebox{0.8\textwidth}{!}{\includegraphics{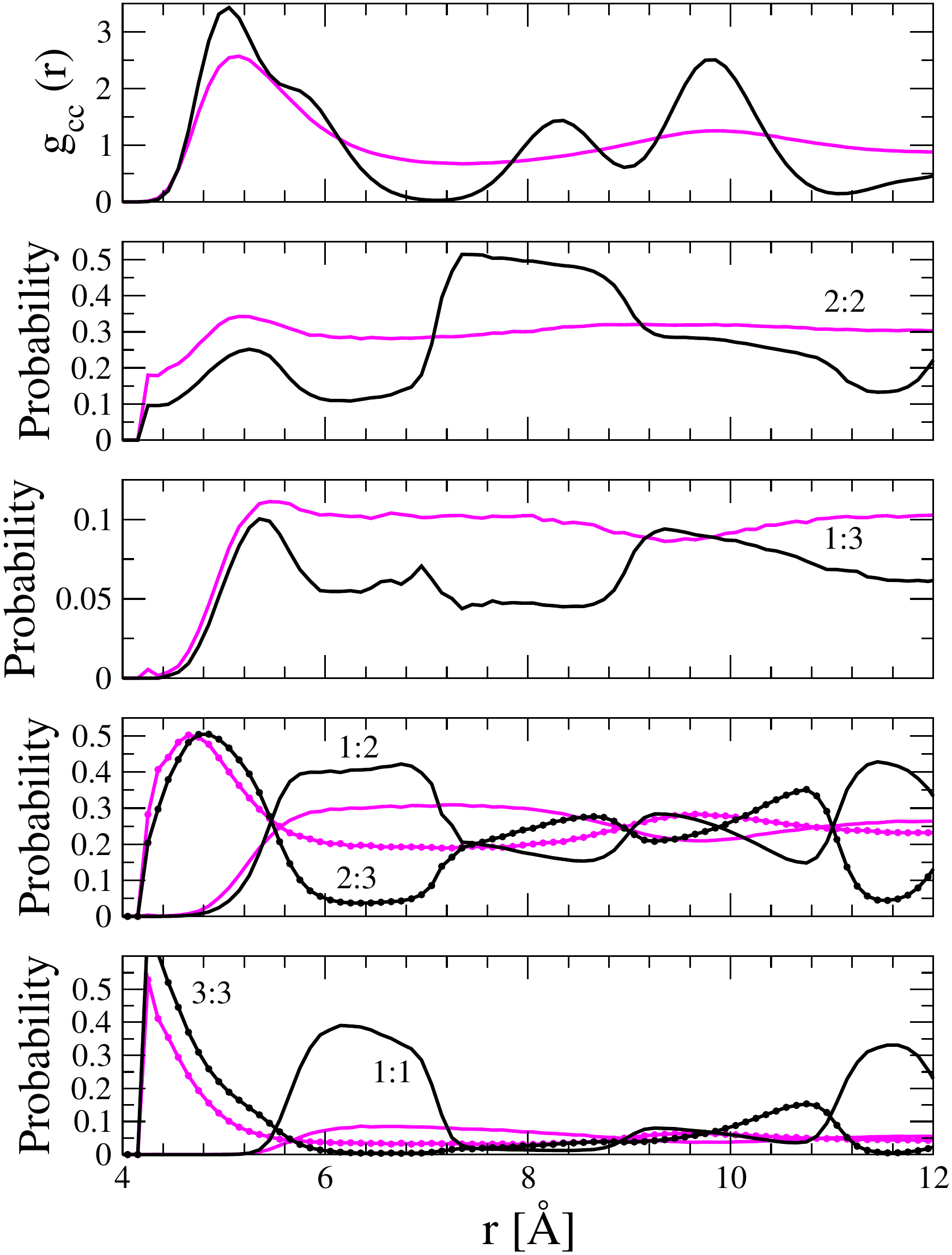}}}
  \caption{\label{fig:sf6_rey_xtal} Comparing orientational correlation probabilities (lower 4 subfigures) in the plastic crystal (190 K, 1 bar, magenta lines) and liquid (225 K, 10.1 bar, black lines) states of SF$_6$ using the classification method of Rey\cite{rey_2007} (extended here for other Platonic shapes) and the molecule centre-centre radial distribution function (top). In the bottom two figures, the groups of 1:1 and 1:2 represented by straight lines, while for 3:3 and 2:3 by straight lines with circles.}
 \end{center}
\end{figure}

One of the advantages of the formalism just introduced is that it can be applied to not just disordered, but crystalline phases, as well. In the body-centered cubic plastic crystalline state at 190 K, each sulfur atom has a definite equilibrium position, while fluorine atomic positions are disordered. This phenomenon implies that although equilibrium positions of sulfur atoms keep the translational symmetry, the molecular centre-centre RDF (FIGURE \ref{fig:sf6_rey_xtal}) expresses noticeable thermal displacements.
Thus, the nearest neighbours in the <111> (or half of cubic diagonal), and second neighbours in the <100> (or side of the cube) directions of the crystal are overlapping in the radial representation, as their ideal values are close: 5.10~\AA{} and 5.89~\AA{}, respectively. Although the present model is probably more disordered than in reality (as discussed in SI. 4.), 
the overlapping behaviour was also observed earlier by Dove and Pawley\cite{dove_1983} using another forcefield. Still, orientational correlations reveal some further features, allowing to separate the contributions of overlapping neighbours: in the <111> direction the 3:3 and 2:3 orientations dominate, meanwhile the 1:3 behaves like in the liquid states and correlations of 2:2 are less abundant. It is interesting to note that the probability of the 3:3 arrangements for close contact is high, while the 2:3 probability is more competitive with 3:3 for the studied liquids and gaseous states. At larger distances, the probabilities of 3:3 and 2:3 drop to the level of random orientations or even less, which shows the approximate limit between the first and second neighbours around 5.5{ }\AA{}. This distance was the rough upper limit of the effect of close contact in the disordered phases. Nevertheless, the most significant characteristic of neighbours in the <100> directions is the broad maxima of 1:1 and 1:2 arrangements. In the <110> (or diagonal of a face of the cube) direction (between 7 and 9 {}\AA) the 2:2 and 2:3 arrangements have maxima. Combining the arrangements found like the maxima of 1:1 to <100> and 3:3 or 2:3 to the <111> directions, the preferred position of fluorine atoms with respect to the lattice can be derived. For the first neighbour, 3 fluorines of each molecule should be placed inside the orientational cones around the cubic body diagonals, while at the same time only one fluorine is allowed to be inside the cone around the side of the cubic cell. As the cones with $\gamma=70.53^o$ half apex angle around the body diagonal (inclined $54.74^o$ to the side of the cubes in the ideal case) cover directions of the three cubic sides, the fluorine atoms should be placed approximately toward the <100> direction from the sulfur positions with higher probability, which agrees well with earlier suggestions\cite{dolling_1979, dove_1984, dove_2002}.

\subsection{Short-range order orientation correlations in models of room temperature C$_{60}$}

As a next step, the orientational correlation scheme is applied to \emph{quantitatively} describe orientational correlations of neighbouring molecules in C$_{60}$, as represented by three forcefields ('Girifalco'\cite{girifalco_1992}, 'Lu'\cite{lu_1992} and 'Sprik'\cite{sprik_1992b}) that had been proven to produce different kinds of short-range order. 

\begin{figure}[htp]
 \begin{center}
  \rotatebox{0}{\resizebox{0.8\textwidth}{!}{\includegraphics{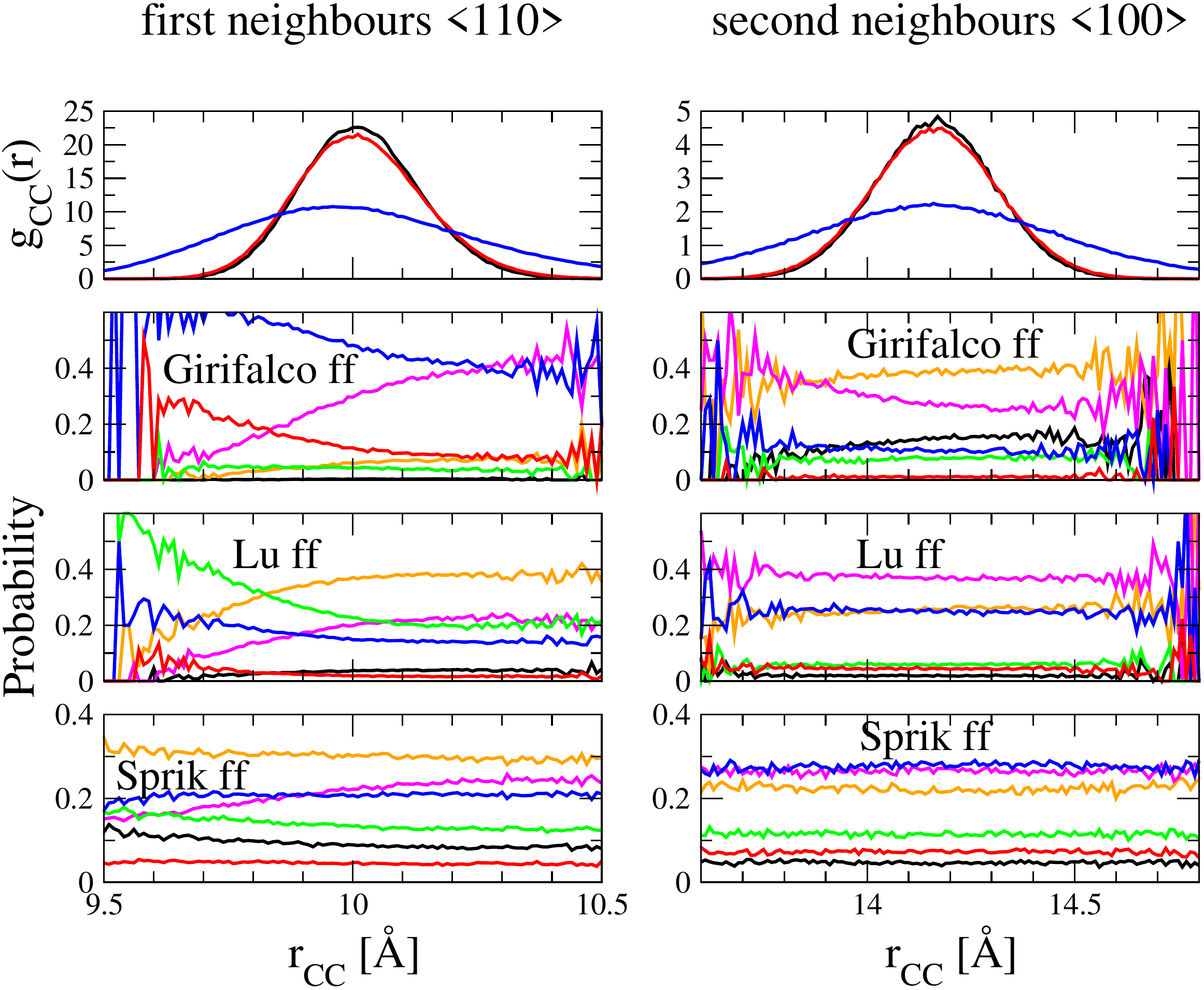}}}
  \caption{\label{fig:c60_rey} Molecular centre-centre radial distribution function (uppermost panels) and orientational correlation probabilities (lower panels) using the classification method of Rey\cite{rey_2007} (extended here for other Platonic shapes) for different MD forcefields for the first (left panels) and second (right panels) nearest neighbours in room temperature C$_{60}$. The examined forcefields are from top to bottom in the lower panels (and color of the lines on the uppermost panels): 'Girifalco'\cite{girifalco_1992} (straight black lines), 'Lu'\cite{lu_1992} (straight red lines) and 'Sprik'\cite{sprik_1992b} (straight blue lines), respectively. The colors of different orientational classes on lower panels are arrangements of face-face (3:3, 'HH'): red lines; corner-corner (1:1, 'PP'): straight black lines; edge-face (2:3, 'HD'): blue line; corner-edge (1:2, 'PD'): orange line; corner-face (1:3, 'PH'): green line; edge-edge (2:2, 'DD'): magenta line.}
 \end{center}
\end{figure}

First, we start the examination of the molecular centre-centre RDF (FIGURE \ref{fig:c60_rey}).
Although some models show a larger centre of mass motion, the most important feature of these RDFs, that the first and second neighbours are distinct. This allows separating orientational correlations unambiguously belonging to the different spheres, in contrast to the plastic crystal phase of SF$_6$, where first and second shells overlap. Touching the broadening issue, it is the result of a different kind of modelling of the intramolecular structure. It was flexible for the 'Girifalco' and 'Lu' models, so the motion of each individual carbon atom is successfully equilibrated by the internal freedom, making the centre less intact from the changes. On the other hand, the 'Sprik' model was rigid, which resulted in a larger displacement of the centres from the equilibrium positions.

Focusing on the orientational correlations, the number of virtual corner atoms in the coordination cone should be assigned to the orientation of C$_{60}$ molecule. One corner atom (the "corner") in the coordination cone means the molecule is oriented by a pentagonal face ('P'), while two atoms ("edge") means between the centres of pentagons: the 'double-bond' ('D'). Finally, the "face" coordination realizes an orientation through the hexagonal faces ('H') of the molecule.

As earlier investigations focused on the short-range order behaviour, here also the first and second neighbours are examined. In order to quantify orientational correlations within a specific coordination shell, the probabilities of each group have been averaged over the given shell. The averaging has been made between the two limiting distances where g$_{cc}$(r) intensity is 1/10 of the maximum intensity within the corresponding shell. Then the shell-averaged probabilities are divided by the corresponding random orientation probability (TABLE \ref{tab:reygrprob}) for icosahedral molecules:
\begin{equation}
 C_{i:j} = \frac{< p_{i:j}(r) >_{shell}}{p_{i:j}^{random}} 
\end{equation}
Thus, its value over 100\% indicates a preferred, while lower refers to a disliked orientation in TABLE \ref{tab:c60_rey_sums}.
\begin{table}[ht]
\caption{\label{tab:c60_rey_sums}Coordination shell averaged probability of orientation groups normalized by the corresponding random orientation probability of room temperature C$_{60}$ by different forcefield models. The 'D', 'H', 'P' are refer to double-bond, hexagonal face, pentagonal face, respectively.}
\resizebox{\textwidth}{!}{
\begin{tabular}[c]{cccccccc}
\hline
Model & Coordination & \multicolumn{6}{c}{Group probability in shell normalized by random orientation [\%]} \\
 & & 1:1 & 3:3 & 1:2 & 2:3 & 2:2 & 1:3 \\
 & shell & 'PP' & 'HH' & 'PD' & 'HD' & 'DD' & 'PH' \\
\hline
'Girifalco' & 1\textsuperscript{st} & 4 & 191 & 23 & 193 & 120 & 33 \\ 
 & 2\textsuperscript{nd} & 221 & 17 & 153 & 43 & 116 & 60 \\ 
'Lu' & 1\textsuperscript{st} & 55 & 34 & 142 & 63 & 78 & 189 \\ 
 & 2\textsuperscript{nd} & 30 & 71 & 102 & 100 & 151 & 47 \\ 
 'Sprik' & 1\textsuperscript{st} & 147 & 72 & 121 & 83 & 87 & 109 \\ 
 & 2\textsuperscript{nd} & 74 & 113 & 90 & 110 & 109 & 91 \\ 
 \hline
\end{tabular}
}
\end{table}
 However, the probability of some arrangements changes by the centre-centre distance within the shell, so values of TABLE \ref{tab:reygrprob} are discussed together with tendencies on FIGURE \ref{fig:c60_rey}.

In the first coordination shell, which corresponds to the closest atoms to <110> directions in the crystalline system, the 'Girifalco' model shows a strong preference of 'HH', 'HD' and a slight preference of 'DD' orientational groups. This suggests that local arrangements try to avoid turning into any of the pentagonal faces of them and prefers hexagonal face - hexagonal face or double-bonds at closer and double-bond-double-bond orientation at distant centre-centre distances, which gives a hint about the rotation mechanism. Looking at the second coordination shell (<100> direction), the opposite 'PP' and 'PD' are favoured strongly by increasing centre-centre distances and the 'DD' slightly, but mainly at closer distances. Even though the preference of 'HD' and unlike of 'PP' and 'PD' found in accordance with earlier study\cite{chaplot_2001}, it contradicts to single-crystal investigation\cite{schiebel_1996} concerning the preference of pentagonal over hexagonal faces for nearest neighbours.

While the pentagonal coordination was disfavored in the first coordination shell by the 'Girifalco' model, the result by the 'Lu' forcefield shows a clear preference of it, agreeing with \cite{schiebel_1996}: both 'PH' (mainly closer distances) and 'PD' (mainly higher distances) kinds of correlations are favoured and the 'PH' arrangement at short distance is more probable. The second shell shows less varied arrangements: only the 'DD' arrangement is preferred. This result is in accord with earlier investigation\cite{chaplot_2001} about the first shell of 'PD' liked and 'PP' disliked.

Interestingly, the 'Sprik' forcefield shows a somewhat opposite kind of preference in comparison with the 'Girifalco', but the deviations from random values are moderate. The pentagonal face related correlations are preferred in the first shell in order of decreasing probability as 'PP', 'PD', 'PH', while in the second shell the hexagonal face related to 'HH', 'HD' and 'DD' correlations are the preferred ones. The 'PP' and 'PD' preference of this model is in agreement with an ab-initio study\cite{tournus_2005}.

\section{Conclusions}\label{conclusions}

Based on the results the following statements can be made:

\begin{enumerate}[(i)]
\item The orientational correlation scheme introduced originally by Rey\cite{rey_2007} is extended to be capable of describing correlations between high symmetry molecules. The orientations of pairs of molecules are classified by the number of corner (aka ligand) atoms for each part found within orientational cones. The axes of the orientation cones correspond to the centre-centre connecting line with the apex at the centres. The extension is made by choosing its apex angle so that the number of corner atoms spans only from 1 to 3 and the average number over random orientation should be 2. Supposing even distribution of atoms, calculation methods for the half apex angle of the cones and asymptotic probabilities of each class are provided.

\item In order to demonstrate the applicability of this method in various condensed phases, the octahedral-shaped SF$_6$ molecule is selected in its gaseous, supercritical fluid, triple point liquid and plastic crystalline phase. The modelling performed by the '7Sites' forcefield\cite{dellis_2010} provided the best agreement with density and total scattering diffraction data.

\item The detailed analysis of different orientational groups for SF$_{6}$ revealed a close contact region in the first coordination shell. It shows a specific orientation pattern of orientational arrangement by distance, similarly to that found for tetrahedral molecules. For non-crystalline states, a medium-range ordered region is found that gradually extends with increasing density, even up to the 4th shell, as observed in high-density states. Its main characteristic feature is that the molecular centre-centre RDF is in phase with the probabilities of arrangements having more corner atoms within the orientational cones than the average 4. Although medium-range order appears for the tetrahedral CCl$_4$ too, it is different from the one observed here. In the plastic crystalline state, close contacts of 3:3, 2:3 correlations are found in the <111> directions, while the second neighbour in <100> showed the preference of 1:1 and 1:2 arrangements. This agrees with earlier studies, that fluorine atoms preferentially oriented in the <100> direction.

\item For the room temperature crystalline C$_{60}$, three models ('Girifalco'\cite{girifalco_1992}, 'Lu'\cite{lu_1992} and 'Sprik'\cite{sprik_1992b}) were tested in the framework of icosahedral symmetry, providing different short-range order. The probability of each arrangement is averaged over the first and the second coordination shells and divided by the corresponding random probability. Thus a quantitative description of orientational correlations is provided for each model, where agreement found in each case with earlier studies. 

\end{enumerate}

It is important to note that this method might be used not only to describe mutual orientational correlations between molecules belonging to the `Platonic` solids (the un-mentioned cube and dodecahedron are the corner-face duals of octahedron and icosahedron, respectively) but also between their distorted forms or any object, as long as the three conditions in section \ref{rey_general} are fulfilled (like for trigonal planar or bipyramidal molecules, where $N_{corner}$ is 3 and 5, respectively). Also, it is possible to use the new algorithm for their mixtures (like tetrahedral and octahedral).


\section*{Acknowledgement}
The author would like to thank the help of L\'aszl\'o Pusztai for discussions. The author is grateful that this project was supported by the János Bolyai Research Scholarship of the Hungarian Academy of Sciences and the National Research, Development and Innovation Office(NKFIH), under grant no. KH130425.



\setcounter{equation}{0}
\renewcommand{\theequation}{S\arabic{equation}}

\setcounter{figure}{0}
\renewcommand{\thefigure}{S\arabic{figure}}
\setcounter{table}{0}
\renewcommand{\thetable}{S\arabic{table}}

\section{Supplementary Information}

\subsection{\label{appendix_random}Calculating the random $P_i$ probabilities}

In order to determine the $P_i$ values for a given molecule, we should consider that each corner atoms are spaced more or less regularly. Thus each $P_i$ probability corresponds to the surface ratio on a unit sphere, where the axis might pass through and there are $i$ number of atoms \emph{inside the cone} (according to EQ. A2 of REF \cite{rey_2007}).

\begin{figure}[ht]
 \begin{center}
  \rotatebox{0}{\resizebox{0.5\textwidth}{!}{\includegraphics[trim={3cm 15.5cm 5cm 1cm}, clip]{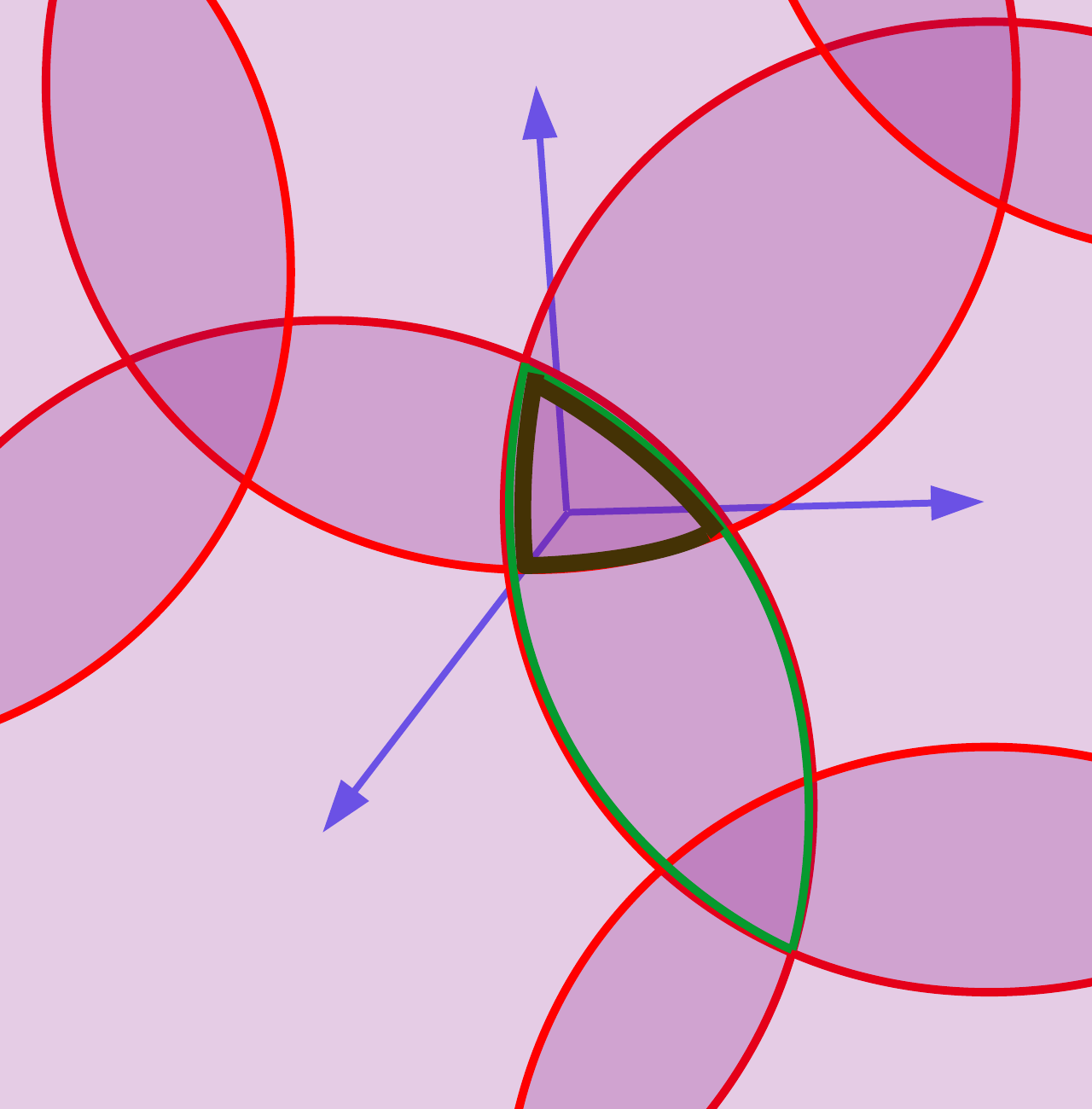}}}
  \caption{\label{fig:surface_map} Surface mapping on the unit sphere. The blue coloured arrows point toward ligand (corner) atoms from the centre of the molecule serving as axes of cones, which intersects the surface at lines marked by red circles. The light, middle and dark purple regions are surface areas containing 1, 2 and 3 ligands if the molecular centre-centre connecting line pass through them. To calculate surface ratios easy, individual surfaces of corner ($p_{cI}$; inside red circle), of edge ($p_{eI}$; surface limited by green lines) and of face ($p_{fI}$; surface limited by black lines) are introduced.}
 \end{center}
\end{figure}

Now, let's fix a molecule and create a cone with $\gamma$ half apex angle around each centre to corner atom axis. By doing this, we get a picture like in fig \ref{fig:surface_map}
around three corner atoms. If a molecular centre-centre connecting line passes through one of these regions on a unit spheres, which might belong to 3, 2 or only 1 such cone(s), it will find the corresponding number of corner atoms inside the orientation cone. Thus, summarizing all of those individual surface elements, which belong to $i$ number of cones, we get the corresponding probability. 

For practical reasons, we introduce the corner ($p_c$: surface ratio of a cone), the edge ($p_e$: surface ratio of 2 overlapping cones) and the face ($p_f$: surface ratio of three overlapping cones) surface ratios composed by the sum of each individual ratios ($p_x=\sum p_{xI}$) on the sphere (see fig \ref{fig:surface_map}). It is important to note, that $p_c$ contains areas, which belong to not just one cone and similarly $p_e$ contain areas, which belong to 3 cones. As a result, $p_c$ and $p_e$ are more than one.

Connecting these ratios to the $P_i$ probabilities, the $P_3$ is exactly equal to $p_f$. On the other hand, the surface of $P_3$ is considered 3 times in the calculations of each $p_c$ and $p_e$. Similarly, the surface of $P_2$ is also considered twice in $p_c$ calculation:
\begin{eqnarray}
 P_3 &=& p_f \\
 P_2 &=& p_e - 3 P_3 = p_e - 3 p_f\\
 P_1 &=& p_c - 2 P_2 - 3 P_3 = p_c - 2 p_e + 3 p_f
\end{eqnarray}
Using EQ. 3 
we get
\begin{eqnarray}
 p_c &=& 2 \label{eq:pcpe}\\
 P_1 &=& P_3 = p_f = p_e-1 \label{eq:P1pe}\\
 P_2 &=& 1 - 2p_f = 3-2p_e \label{eq:P2pe}
\end{eqnarray}
EQ. \ref{eq:pcpe} is just the reformulation of EQ. 1, 
that the orientation cone should contain 2 corner atoms. The least 2 equation states, that either $p_f$ or $p_e$ should be calculated to determine all $P_i$ probabilities. Although, appendix of\cite{rey_2007} calculated the $p_{fI}$ for tetrahedral molecules, in the following the $p_{eI}$ is calculated. It takes into account only pair of corner atoms, whose associated parameter (the corner-centre-corner bond angle) is more convenient to use in the case of non-regular shaped objects.

\begin{figure}[ht]
 \begin{center}
  \rotatebox{0}{\resizebox{0.5\textwidth}{!}{\includegraphics[trim={0 18cm 3cm 0}, clip]{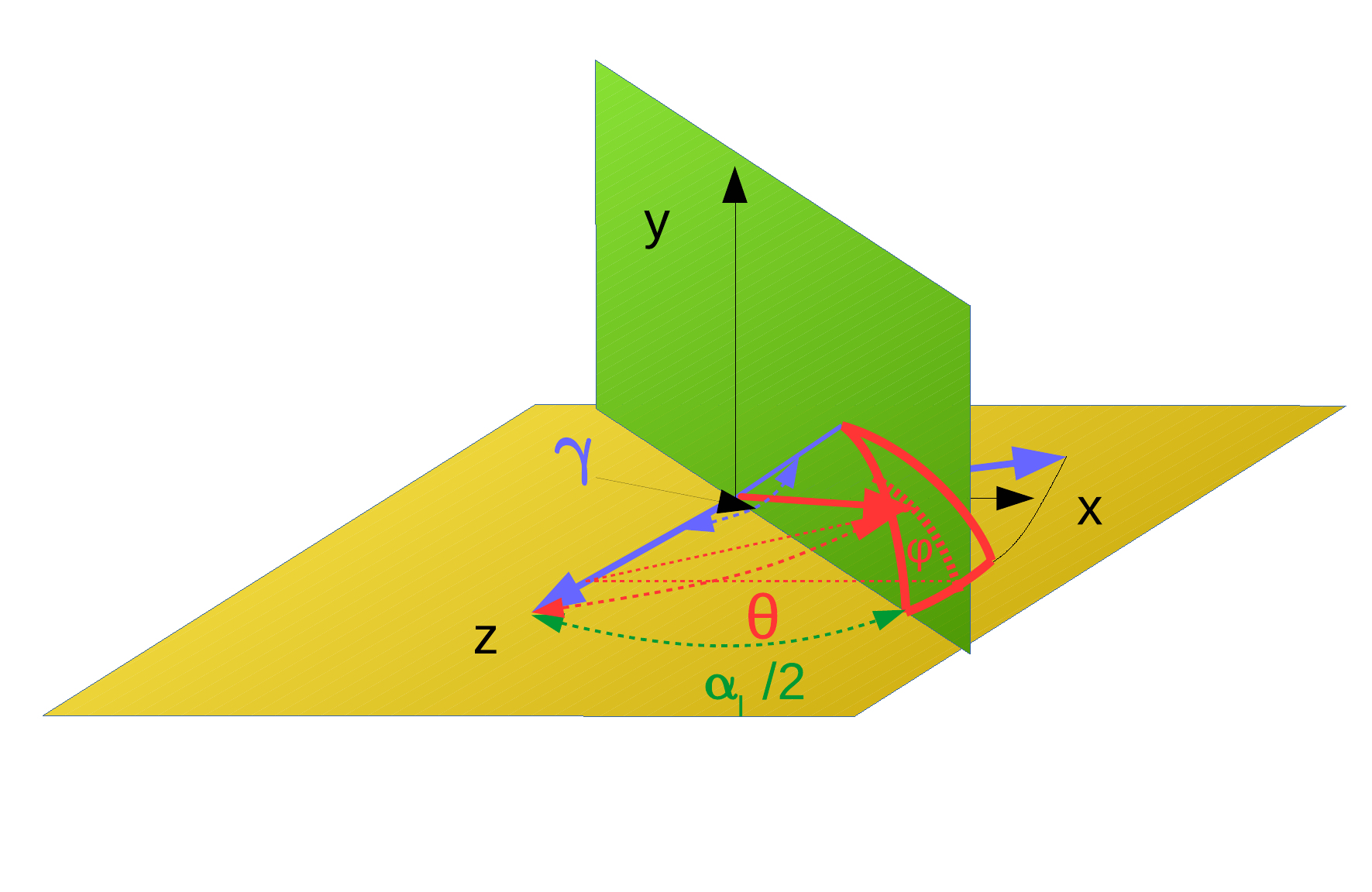}}}
  \caption{\label{fig:integration} Calculation of 1/4 of the surface integral belonging to $p_{eI}$. The surface to be calculated is limited by thick red lines, the blue colored arrows are the axis of the cones, whereas each plane is a symmetry plane.}
 \end{center}
\end{figure}

In order to calculate the surface ratio, we should consider two cones with a common apex, both having the same half apex angle ($\gamma$) and their axes are separated by $\alpha_I$ degrees from each other. The surface ratio will be the common area within the two cones on the unit sphere in comparison with the whole surface. Let's choose a Cartesian coordinate system, direct the 'z' axis to the axis of one of the cones, place 'x' axis in the plane determined by the axes of each cone and 'y' perpendicular to them (see FIG. \ref{fig:integration}).
The surface to be calculated posses two planes of symmetry: one is the 'xz' plane at y=0 and the other is the bisecting plane inclined by $\alpha_I/2$ degrees to the z-axis ($\tan{\alpha_I /2}=x/z$ on the plane). Taking that small surface element on the unit sphere ($S_I$), which is limited by the two symmetry planes and the intersecting cone, the corresponding probability:
\begin{equation}
 p_{eI}=\frac{4S_I}{4\pi}=\frac{S_I}{\pi}.
\end{equation}

In order to calculate $S_I$, we parametrize its surface to spherical coordinates. Applying the Cartesian-spherical conversion form on unit of spheres $(x,y,z)=(\sin{\theta}\cos{\phi},$ $\sin{\theta}\sin{\phi}, \cos{\theta})$, the limits are: 
\begin{eqnarray}
 0 \leq \phi &\leq& \arccos{\left\{ \frac{\tan{\alpha_I/2}}{\tan{\theta}}\right \}}, \\
 \alpha_I/2 \leq \theta &\leq& \gamma
\end{eqnarray}
Thus, the individual ratio is obtained by evaluating the following surface integral:
\begin{eqnarray}
 p_{eI}&=&\frac{S_I}{\pi}=\frac{\int \limits_{\alpha_I/2}^{\gamma} \int \limits_0^{\arccos{\left\{ \frac{\tan{\alpha_I/2}}{\tan{\theta}}\right \}}} d \phi \sin{\theta} d \theta }{\pi}=\nonumber \\ &=&\frac{\arctan{\sqrt{\left( \frac{ \sin{\gamma}}{\sin{\alpha_I/2}} \right)^2-1}} - \cos{\gamma}\arccos{\left(\frac{\tan{\alpha_I/2}}{\tan{\gamma}}\right)}}{\pi}
\end{eqnarray}

Introducing the $A=\cos{\alpha_I}$ and $G=\cos{\gamma_I}$ notations, the individual surface ratio of the edge becomes
\begin{equation}
 p_{eI}=\frac{1}{\pi}\left\{\arctan{\sqrt{\frac{1+A-2G^2}{1-A}}}-G\arccos{\left(\sqrt{\frac{1-A}{1+A}}\frac{G}{\sqrt{1-G^2}}\right)}\right\}
\end{equation}

Evaluating it to tetrahedral ($A=-1/3$, $G=0$), octahedral ($A=0$, $G=1/3$) and icosahedral ($A=1/\sqrt{5}$, $G=2/3$) symmetries, we get:
\begin{eqnarray}
 \tilde{p}_{eI\textrm{T}}&=&\frac{1}{\pi}\arctan{\frac{1}{\sqrt{2}}} \\
 \tilde{p}_{eI\textrm{O}}&=&\frac{1}{\pi}\left\{\arctan{\left(\frac{\sqrt{7}}{3}\right)}-\frac{1}{3}\arccos{\left(\frac{1}{2\sqrt{2}}\right)}\right\} \\
 \tilde{p}_{eI\textrm{I}}&=&\frac{1}{\pi}\left\{\arctan{\sqrt{\frac{7+5\sqrt{5}}{18}}}-\frac{2}{3}\arccos{\left(\sqrt{\frac{2(3-\sqrt{5})}{5}}\right)}\right\}
\end{eqnarray}

The individual surface ratios should be summed up in general, or simply multiplied with the corresponding number of edges in regular (aka Platonic solids) case to obtain $p_e$, then applying EQ. \ref{eq:P1pe} and \ref{eq:P2pe} to get any of the $P_i$ random orientation probabilities (TABLE \ref{tab:rorientprob}).
\begin{table}[ht]
\caption{\label{tab:rorientprob}Numerical values of random orientation probabilities for some high-symmetry molecules.}
\resizebox{\textwidth}{!}{
\begin{tabular}[c]{ccccccccc}
\hline
symmetry & $N_{corner}$ & $N_{edges}$ & $\cos{\alpha}$/$\alpha$ [$^o$] & $\cos{\gamma}$/$\gamma$ [$^o$] & $p_{eI}$ & $p_e$ & $P_1=P_3=p_f$ & $P_2$ \\
\hline
tetrahedral & 4 & 6 & $-\frac{1}{3}$/109.47 & 0/90 & 0.19591 & 1.17548 & 0.17548 & 0.64904 \\
octahedral & 6 & 12 & 0/90 & $\frac{1}{3}$/70.53 & 0.10173 & 1.22075 & 0.22075 & 0.55850 \\
icosahedral & 12 & 30 & $\frac{1}{\sqrt{5}}$/63.43 & $\frac{2}{3}$/48.19 & 0.04175 & 1.25251 & 0.25251 & 0.49498 \\
\hline
\end{tabular}
}
\end{table}

Note the exact agreement with the calculated value of $P_1$ of Rey\cite{rey_2007} for tetrahedral molecules. 

\subsection{\label{appendix_md}Details of the performed molecular dynamics simulations}

For sulfur hexafluoride, calculations at various thermodynamic conditions done, where published diffraction datasets\cite{strauss_1994, dove_2002, rmcprofile_tutorial, powles_1983} were available. Thus, both NpT and NVT simulations have been performed in the supercritical state (temperature of 398 K and pressures of 128 bar, 155 bar, 394 bar and 1827 bar\cite{strauss_1994}), in the orientationally disordered/plastic crystalline phase I (190 K and 1 bar\cite{dove_2002, rmcprofile_tutorial}), in the liquid state close to the triple point (225 K and 10.0552 bar -- no measured diffraction data are available here). Only NVT calculations have been performed in the gaseous phase (293.15 K and 10.2 bar\cite{powles_1983}). Initial configurations containing 5000 molecules are created for gaseous, liquid and fluid phases by randomly placing and orienting molecules into the simulation box determined by the density. In the crystalline bcc phase, each of the 5488 molecules (cell size 14x14x14 of the crystallographic unit cell with 5.89 {}\AA{} lattice constant) are oriented randomly around equilibrium positions of sulfur.

For fullerene, only its plastic crystalline phase is modelled at room temperature by NVT simulations using three forcefields (FIG. \ref{tab:c60ffpars}). Each simulation started from an 8x8x8 fcc unit cell (corresponds to 2048 molecules, with lattice constant of 14.16 {}\AA) points as centres of gravity of C$_{60}$, where molecules randomly oriented.

\begin{table}[ht]
\caption{\label{tab:c60ffpars}Lennard-Jones parameters and partial charges applied for fullerene MD simulations at room temperature, ambient conditions. For flexible molecules (applied with 'Girifalco'\cite{girifalco_1992} and 'Lu'\cite{lu_1992} forcefields), the flexible intramolecular potential set of Monticelli et al.\cite{monticelli_2012} has been used for carbon atoms: bondlengths of 0.14 and 0.145 nm with 392459.2 $\frac{kJ}{mol*nm^2}$, bond angles of 120 and 108 degrees with 527.184 $\frac{kJ}{mol*rad^2}$, improper dihedrals of 143 degrees with 100 $\frac{kJ}{mol*rad^2}$ force constant.}
\resizebox{\textwidth}{!}{
\begin{tabular}[c]{ccccccccc}
\hline
\multicolumn{2}{c}{Forcefield} & \multicolumn{3}{c}{Carbon atoms} & \multicolumn{3}{c}{Double bonds} & Single bond\\
name & reference & charge [$e$] & $\sigma$ [nm] & $\epsilon$ [$\frac{kJ}{mol}$] & charge [$e$] & $\sigma$ [nm] & $\epsilon$ [$\frac{kJ}{mol}$] & charge [$e$] \\
\hline
'Girifalco' & \cite{girifalco_1992} & 0 & 0.346903 & 0.276427 &  &  &  & \\
'Lu' & \cite{lu_1992} & 0 & 0.3407 & 0.28598 & -0.54 &  &  & 0.27 \\
'Sprik' & \cite{sprik_1992b} & 0.175 & 0.34 & 0.1247 & -0.35 & 0.36 & 0.1247 & \\
\hline
\end{tabular}
}
\end{table}

During the simulations, the Verlet cut-off scheme has been used with van der Waals cut-off at 2.0 nm (1.0 nm for C$_{60}$). For electrostatics, the particle-mesh Ewald algorithm has been applied, with the same Coulombic cut-off distance when partial charges were non-neutral. The simulation time step was 1 fs. The configuration energies were minimized first, then equilibrated at different levels: for 0.5 ns Berendsen-thermostat ($\tau_T$ = 0.2 ps, for C$_{60}$ it was 2 ps) applied at the right temperature first, then Nose-Hoover thermostat for 2 ns. For NpT runs, an additional 200 ps by Berendsen barostat (with $\tau_p$= 1.0) and at least 1 ns by Parinello-Rahman barostat were applied. Carrying out equilibration, production runs were started for 10 ns, saving coordinates at each 100 ps.

In the case of simulations with the 'Lu' and 'Sprik' forcefields of C$_{60}$, a modelling sequence at 700K, then at 500K and finally at 300K has been conducted, where each modelling started from the final configuration of the previous simulation.

For density determination NpT, while for other cases NVT simulations are performed.

\subsection{\label{appendix_totscatcalc}Total scattering powder diffraction patterns: available data and way of calculation from simulations}

All measured total scattering powder diffraction pattern have been taken from literature\cite{powles_1983, dove_2002, rmcprofile_tutorial, strauss_1994, leclercq_1993}. Numerical experimental datasets were available for the gaseous\cite{powles_1983} and plastic crystalline phase of SF$_6$ at 190K\cite{dove_2002, rmcprofile_tutorial} while for supercritical fluid states\cite{strauss_1994} and C$_{60}$ at room temperature\cite{leclercq_1993} datasets have been digitized from figures of differential cross-sections and intensity, respectively. 

For crystalline phases, total scattering structure factors have been calculated using a new implementation of the RMCPOW method\cite{rmcpow_1999}. For SF$_6$ 11 configurations were used for calculation, separated by at least 1 ns of NVT runs, while only the final configuration was taken for C$_{60}$s.  Bragg contributions convoluted by the instrumental resolution function, which was "TOF profile function-2" of GSAS\cite{gsas_2004} for SF$_6$ and Gaussian shape for C$_{60}$. Their parameters are optimized by a trial configuration. Finally, individually calculated patterns are averaged.

For other phases, the 'rdf' program of GROMACS provided the partial RDFs, averaged over 101 configurations taken at each 100 ps. Then RDFs are Fourier-transformed and weighted sum created according to neutron coherent scattering lengths and composition.

Generally, during the comparison of the measured and simulated total structure factors, the former is rescaled and a constant background is subtracted to avoid errors from different normalization. For crystalline datasets, the measured pattern rescaled and corrected by 2nd order polynomials above 2.5 {}\AA$^{-1}$ first, then it has been compared directly over the full measured range.

In order to compare the calculated pattern ($I^{calc}$) with the corrected measured one ($I^{meas}$), a reliability factor analogous to $R_{wp}$ at Rietveld-refinement\cite{young_1993} is used:
\begin{equation}\label{eq:reliability}
 R_{wp} = \sqrt{\frac{\sum_i (I^{meas}_i - I^{calc}_i)^2}{\sum_i (I^{calc}_i)^2}}
\end{equation}
The difference is the weights by points are equal and the denominator contains the calculated data instead of the measured one.

\subsection{\label{appendix_ffcheck}Checking the performance of SF$_6$ forcefields against total scattering powder diffraction data and density}

The agreements between measured\cite{powles_1983, strauss_1994, rmcprofile_tutorial} and NVT MD simulated structure factors are compared in TABLE \ref{tab:sf6rwp} using the reliability factor defined in EQ. \ref{eq:reliability}.
\begin{table}[ht]
\caption{\label{tab:sf6rwp}R$_{wp}$ [\%] of simulated and measured\cite{powles_1983, strauss_1994, rmcprofile_tutorial} total scattering structure factors of SF$_6$ for NVT simulations.}
\begin{center}
\begin{tabular}{ccccccc}
\hline
 Reference & \cite{powles_1983} & \cite{strauss_1994} & \cite{strauss_1994} & \cite{strauss_1994} & \cite{strauss_1994} & \cite{rmcprofile_tutorial}\\
 Tempereture [K] & 293.15 & 398 & 398 & 398 & 398 & 190 \\
 Pressure [bar] & 10.2 & 128 & 155 & 394 & 1827 & 1 \\
 Density [kg/m$^3$] & 71.306 & 850 & 1000 & 1400 & 1850 & 2374 \\
 \hline
 Kinney & 3.59 & 8.09 & 10.29 & 9.84 & 10.25 & 27.9 \\
 Kinney-opt & 3.19 & 9.61 & 9.78 & 11.02 & 11.69 & 27.8 \\
 Pawley & 3.01 & 7.34 & 11.41 & 11.07 & 9.06 & 29.3 \\
 Pawley-opt & 3.28 & 9.11 & 9.77 & 11.10 & 13.03 & 27.3 \\
 Strauss & 4.97 & 34.1 & 41.6 & 34.7 & 38.0 & 32.7 \\
 Strauss-opt & 3.14 & 9.05 & 9.24 & 10.52 & 11.07 & 27.6 \\
 Dellis & 3.15 & 9.19 & 9.91 & 10.89 & 11.67 & 27.3 \\
\hline
\end{tabular}
\end{center}
\end{table}

\begin{figure}[ht]
 \begin{center}
  \rotatebox{0}{\resizebox{0.8\textwidth}{!}{\includegraphics{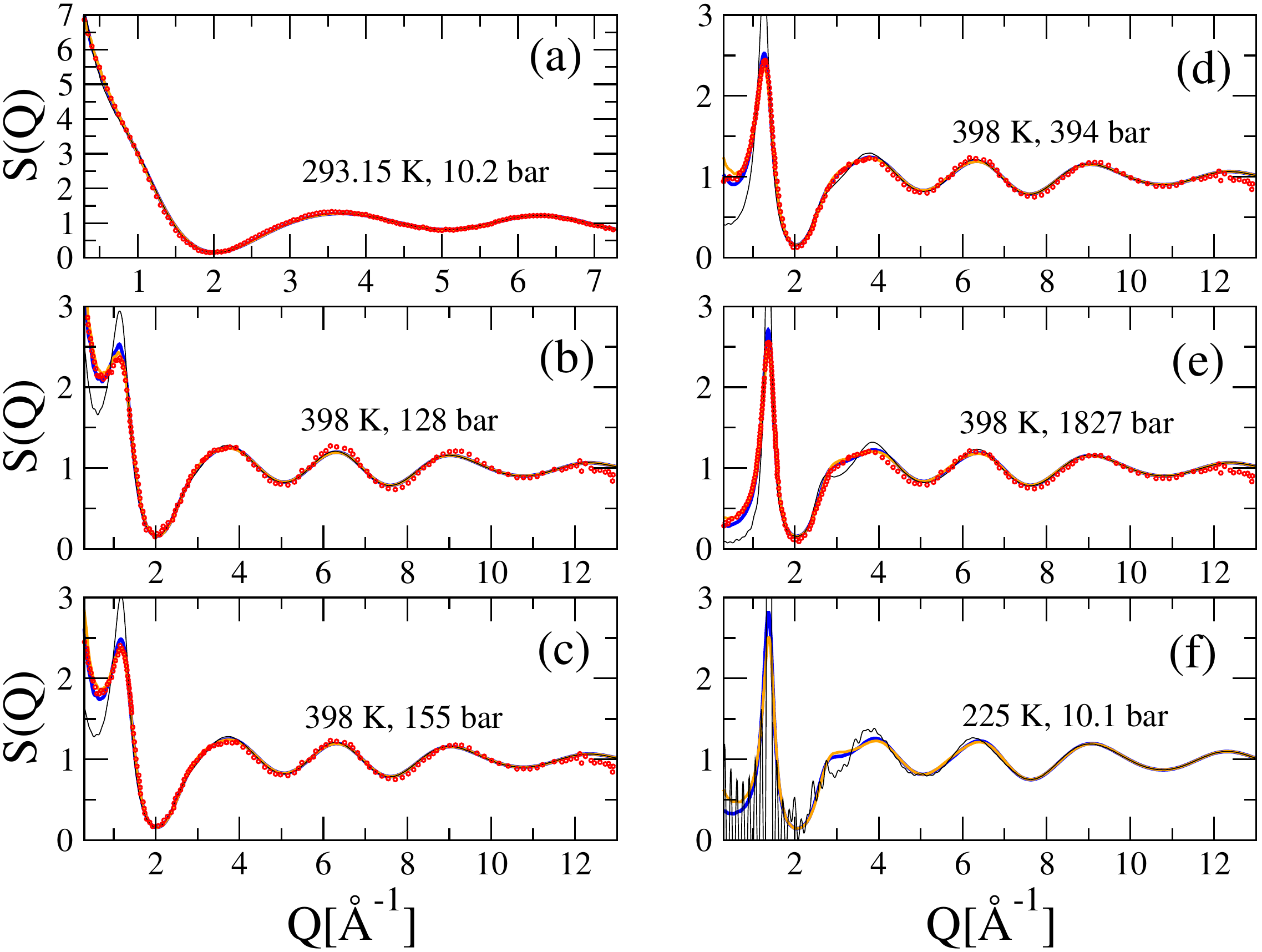}}}
  \caption{\label{fig:sf6_liq} Comparison of the MD simulated and experimental\cite{powles_1983, strauss_1994} total scattering structure factors of gaseous\cite{powles_1983} (a), supercritical fluid\cite{strauss_1994} (b)-(e), and prediction for close to triple point liquid state (f) of SF$_6$. Experimental datasets\cite{powles_1983, strauss_1994}: red circles; MD with 'Dellis' forcefield\cite{dellis_2010}: straight blue lines; with 'Pawley' forcefield\cite{pawley_1981}: straight orange lines; with 'Strauss' forcefield\cite{strauss_1994}: straight black lines.}
 \end{center}
\end{figure}

Comparing the performance of different forcefields, the worst agreements are found for the 'Strauss' one for each condition, while other forcefields show similar agreement. It is interesting to note that the non-optimized 'Pawley' forcefield shows the best agreement for some states. As a consequence, the 'Dellis', 'Pawley' and 'Strauss' forcefields are selected to represent the corresponding agreement in FIGURE \ref{fig:sf6_liq} for the disordered phases.

In the gaseous phase (FIGURE \ref{fig:sf6_liq}a), each forcefield provides almost perfect agreements with the measured dataset\cite{powles_1983} because the intramolecular structure factor dominates the total structure factor. Focusing on more dense states, the 'Strauss' forcefield total structure factors have not reproduced the measured ones, and close to the triple point liquid phase (FIGURE \ref{fig:sf6_liq}f) large oscillations appeared as the simulated system becomes crystalline. An earlier MD investigation\cite{dellis_2010} also showed that this forcefield results in peak shifts of the total radial distribution function in comparison with the experimental one, while other forcefields show similar behaviours to each other.

\begin{figure}[hp]
 \begin{center}
  \rotatebox{0}{\resizebox{0.8\textwidth}{!}{\includegraphics{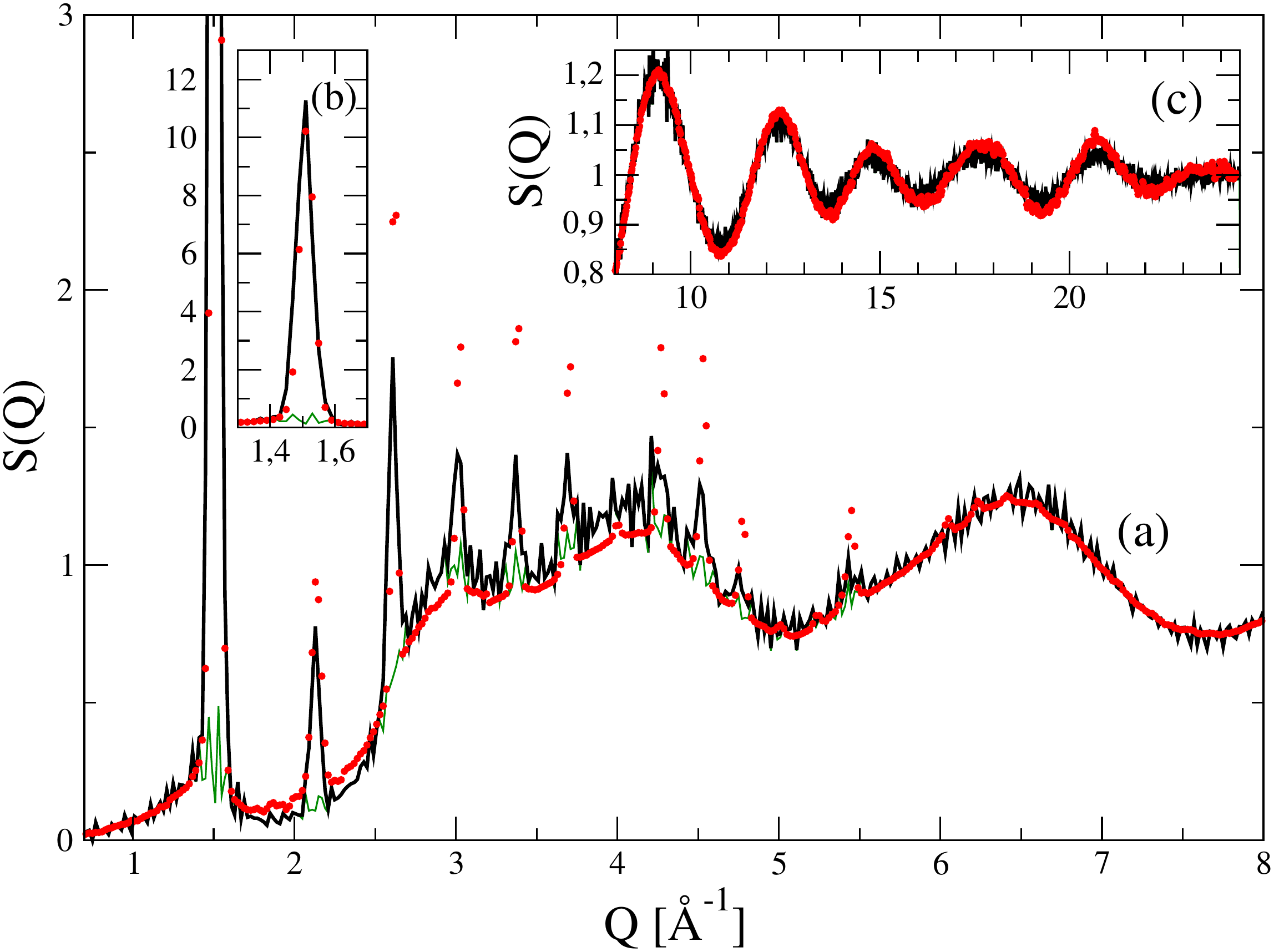}}}
  \caption{\label{fig:sf6_xtal} Comparison of MD simulated (with the 'Dellis' forcefield\cite{dellis_2010}) and experimental\cite{rmcprofile_tutorial} total scattering powder structure factors of SF$_6$ at 190 K in the plastic crystal state. The low-Q-part is shown on (a), while insets show the agreement at (110) Bragg-reflection (b) and high Q (c). Experimental \cite{rmcprofile_tutorial} datasets: red circles; MD simulated total (Bragg+diffuse) and only diffuse scattering contributions: straight black and green lines, respectively.}
 \end{center}
\end{figure}

Observing the performance of the other forcefields, agreements at supercritical states (FIGURES \ref{fig:sf6_liq}b-e) are excellent. Only slight differences appear at the low-Q part: the total structure factors of 'Pawley' are slightly less intense than those of 'Dellis': in FIGURE \ref{fig:sf6_liq}d intensity starts to rise below 0.5{ }\AA $^{-1}$. The other differences between simulated and measured total structure factors can be explained by errors of digitalization. For the plastic crystalline state (FIGURE \ref{fig:sf6_xtal}),
the simulated pattern provides satisfactory agreement with the total structure factor. Three main differences can be identified: the Bragg-peak intensities; diffuse scattering intensity differences between 1.5 and 5{ }\AA $^{-1}$; and intensity differences beyond 15{ }\AA $^{-1}$. The first one might result from MD simulations provided a slightly disordered model in comparison with reality as the simulated intensity of Bragg-peaks become smaller as Q increases. The second one might come that in this range the measured 3 bank datasets overlap, making their alignment slightly difficult. The last one might originate from the limits of harmonic approximation of intramolecular bond lengths and angle bending instead of a more realistic approximation for far from its equilibrium value. Here, the 'Dellis' model performs better in comparison with the 'Pawley'.

The average densities produced by NpT runs are shown in TABLE \ref{sf6dens} for almost every forcefield.
\begin{table}[ht]
\caption{\label{sf6dens}Differences of NpT simulated densities from the experimental one in [kg/m$^3$], using different forcefields for SF$_6$.}
\begin{center}
\begin{tabular}{ccccccc}
\hline
 Reference & \cite{strauss_1994} & \cite{strauss_1994} & \cite{strauss_1994} & \cite{strauss_1994} & \cite{dove_2002, rmcprofile_tutorial} & \cite{funke_2002} \\
 Tempereture [K] & 398 & 398 & 398 & 398 & 190 & 225 \\
 Pressure [bar] & 128 & 155 & 394 & 1827 & 1 & 10.1 \\
 Density [kg/m$^3$] & 850 & 1000 & 1400 & 1850 & 2374 & 1842 \\
 \hline
 Kinney & 114 & 77 & 55 & 91 & -4 & 266 \\
 Kinney-opt & -8 & -36 & -16 & 45 & -69 & 84 \\
 Pawley & -112 & -117 & -12 & 118 & 81 & -185 \\
 Pawley-opt & -17 & -42 & -20 & 34 & -45 & 42 \\
 Strauss-opt & -8 & -34 & -9 & 57 & -43 & 71 \\
 Dellis & 12 & -14 & -3 & 55 & -41 & 58 \\
 \hline
\end{tabular}
\end{center}
\end{table}
It does not contain gaseous phase datasets, because large volume fluctuations prevented performing NpT simulations, while the 'Strauss' forcefield provided a poor agreement to the total structure factors in the NVT ensemble, so the corresponding NpT simulations have not been done. Comparing the experimental with the simulated densities at 6 states, we can conclude that generally, the optimized and the 7 sites 'Dellis' forcefields perform better even for the 190K and 225K states that are below the applied range used for optimization by Dellis and Samios\cite{dellis_2010}. The strong negative difference in the density of the 'Pawley' forcefield for states near the triple point refers to shifts towards the gaseous state (the difference is notable in terms of total scattering structure factor in FIG. \ref{fig:sf6_liq}f), while positive differences for 'Kinney' refer to some transition to the crystalline state.

Taking into account the agreement with experimental density and with neutron total scattering diffraction patterns, the 7 sites 'Dellis' (with partial charges) or the 6 sites 'Pawley-opt' forcefields are recommended to use.

\bibliography{platonic}

\end{document}